\documentclass[reprint,amsmath,amssymb,aps]{revtex4-2}

\usepackage{graphicx}
\usepackage{dcolumn}
\usepackage{bm}
\usepackage{braket}

\begin{document}

\title{The indivisibility of a quantum-corrected AdS black hole with phantom global monopoles}

\author{Tiantong Cheng}

\affiliation{
	School of Data Science, Shanghai Lida University, Shanghai 201608, China}
\author{Hongbo Cheng}
\email{hbcheng@ecust.edu.cn}
\affiliation{
School of Physics, East China University of Science and Technology, Shanghai 200237, China\\
The Shanghia Key Laboratory of Astrophysics, Shanghai 200234, China}

\date{\today}

\begin{abstract}
We study the indivisibility of a quantum-corrected AdS black hole involving two kinds of global monopoles, regular or phantom ones to declare the effects from the quantum fluctuation, monopole factors and the negative cosmological constant on the evolution of black holes. We focus our attention on the possibility that this kind of black holes break into their own two parts because of the second law of thermodynamics. We derive and calculate the entropies of the initial black hole and the broken parts respectively and the entropy difference relates to the black holes structure including the quantum-gravity oscillation, global monopoles and the influence from AdS environment. It is found that the two kinds of global monopoles both keep the black holes intact instead of splitting because the entropy difference is negative. The considerable quantum fluctuation compels the defference to be positive and the fragmentation of the black holes may happen in the case that the mass of one fragmented black hole is tiny and the other one's mass is huge. The AdS radius can increase the whole value of change in entropy and the difference near the midpoint of mass ratio will become positive, so the large enough radius may lead the isolated black hole to become two parts nearly equal in the mass.
\end{abstract}

\maketitle


\section{Introduction}

There have been numerous stars during the evolution of universe and some of the stars have been proceeding the gravitational collapse to become black holes [1]. The black holes as solutions to the Einstein equations theoretically are perceived as perfect absorbers classically, which means that no electromagnetic radiation can escape from the sources [1, 2]. In the process of universe evolution, this kind of compact bodies generated and have been envolving and a lot of significant attentions have been paid for the celestial objects for decades [1, 3-6]. It is significant to conduct the detailed investigations on the models in several directions because the black holes have their own characteristics relating to their structures and environments [1, 2]. There must be many kinds of black holes with different metrics respectively [1, 2].

It is crucial to study the quantum-corrected black holes involving phantom global monopoles to test the general relativity and address the problems at the intersection of Einstein gravity and quantum mechanics further [7]. The expansion of the universe has been proceeding with decreasing temperature. When the temperature dropped to the critical points in the early stage of the universe, the vacuum phase transition occured while various topological defects such as domain walls, cosmic strings and monopoles formed [8, 9]. These topological defects generated due to a breakdown of local or global gauge symmetries theoretically [8, 9]. A global monopole as a spherically symmetric topological defect came from the phase transition of a system composed of a self-coupling triplet of a scalar field whose original global $O(3)$ symmetry is spontaneously broken to $U(1)$ [8, 10]. During the evolution of the universe there should exist a kind of gravitational sources that contain global monopoles and the metric of the sources has a solid deficit angle leading all electromagnetic rays close to the source to be deflected at the same angle [11]. It should be emphasized that the black holes swallowing the global monopoles like the most of other ones also possess the singularities causing the spacetime curvature around the compact objects to be infinity [11]. It is feasible to incorporate the quantum corrections into the black hole metrics to resolve the singularities from the Einstein equations [12]. It should be pointed out that the singularities originated from the general relativity lead the spacetime curvature to be infinity and violate the laws of physics and the quantum theory of gravity can be used to resolve them [7, 12]. It is effective to probe the quantum-corrected black holes by means of various techniques. A lot of efforts have been contributed to the compact stars with quantum effect. The black hole's phenomena such as Hawking radiation, information paradox etc. from the black hole metrics involving the quantum corrections are discussed to test the quantum gravity effects [13-16]. The black hole's metric with quantum corrections can help us to deepen our comprehension for the spacetime quantum structure [17]. The black holes in the cosmological background with a negative cosmological constant viewed as a gauge-gravity duality in the boundary of anti-de Sitter (AdS) spacetime are also popular models named as AdS black holes and it is well known that the AdS spacetime is widely used to quantum field theory and cosmology [18]. Based on the statements above, there probably exist quantum-corrected AdS black holes with global monopoles and it is clear that their metric functions are not simple [7, 11, 19, 20]. It is useful to probe this kind of black holes in various directions. The influence of regular global monopole or phantom one swallowed by the compact source on the motion of neutral particles nearby was investigated [21].

It is significant to scrutinize the effects from quantum correction to the AdS black hole metric in the presence of global monopoles on the thermodynamics of black hole to confirm the compact object's stability. The thermodynamic properties of quantum-corrected AdS black hole with an ordinary or a phantom global monopole were studied [22]. The properties are shown as Hawking temperature and specific heat capacity and also include the geodesics, shadow radius and energy emission rate of the black holes [22]. The configurations such as quantum fluctuation, two kinds of global monopoles can significantly alter the near-horizon geometry in the AdS spacetime to declare the nature of heat capacity and the gravitational bodies with negative heat capacity will evaporate, enriching the understanding of the thermodynamic stability of black holes [22].

It is necessary to discuss the fragmentation instability belonging to the thermodynamics of the spherically symmetric quantum-corrected black hole incorporating an ordinary or a phantom global monopole in an AdS space background. The thermodynamic stability of black holes is of growing interest. The isolated black holes may evaporate to disappear or may also split [1]. Whether the black holes release their energy completely depends on the nature of heat capacity [1]. A lot of efforts have been paid for the researches on the several kinds of AdS black holes [1, 22-28]. The authors of Ref.[22] computed these thermodynamic quantities like Hawking temperature, entropy, specific heat capacity combined with the quantum-corrected parameter and the inclusion of ordinary monopole or phantom ones to declare the differences in results for the two kinds of monopoles respectively. They found that the differences in the effects between the cases of the different monopoles can be distinguished within a range of quantum-corrected factor values and small energy scale parameters [22]. As mentioned above, a lot of attentions have been focused on the nature of the heat capacity of the quantum AdS black holes swallowing regular or phantom global monopoles, the black holes with negative heat capacity evaporating completely. It should be pointed out that the thermodynamic stability of black holes does not just depend on whether the compact objects will evaporate and also on whether the black holes will split due to the sign of entropy difference between the initial state and the final ones. The indivisibility of black holes certainly belongs to their stability. The fragmentation probability may violate the black hole's integrity [29]. The understanding of black hole stability should be deepened because the fragmentation of black hole also disrupts its stability. The evolution of black hole may be thought to be driven by the second law of thermodynamics and whether the evolution could happen depends on the sign of the entropy difference of the black holes in the different stages respectively. If the final entropy consisting of the sum of entropies for broken bits of the original black hole is smaller than the initial one, the negative entropy difference, the isolated black hole will be stable instead of splitting. Conversely, the positive difference means that the compact body can break into parts [29]. The fragmentation scheme was used to investigate the evolution of a serious of black holes such as rotating anti-de Sitter black holes [30], black holes with a Gauss-Bonnet term [31], charged anti-de Sitter black holes [32], black holes with $f(R)$ global monopole under the generalized uncertainty principle [33] and Einstein-Euler-Heisenberg-AdS black holes [34].

It is necessary to probe the integrity or fragmentation instability of a quantum-corrected AdS black hole with an ordinary or a phantom global monopole. The insights into the thermodynamic stability of this kind of black holes need to be enriched. We can not ignore the possibility of fragmentation of these black holes. The black holes structures are complex and their metrics contain more factors. The influences from quantum fluctuation and global monopoles on the thermodynamics of black holes have been studied [22], but to our best knowledge, little contributions have been paid for indivisibility as another kind of thermodynamic stability of this kind of black holes. We plan to investigate whether the quantum fluctuation, two kinds of global monopoles lead the black holes to break into two sections with the help of the technique from Ref.[29]. We derive and calculate the entropy of black hole governed by the factors relating to the quantum gravity, monopoles and the radius of world with negative cosmological constant. We compare the entropy of the initial black hole with that of the final system consisting of two fragmented black holes to find how the black hole entropy changes. In the process of our research on the entropy difference which decide whether the fragmentation will happen under the second law of thermodynamics, we wonder how the quantum fluctuation, ordinary or phantom monopole and the environment affect the final fate of black hole.

\section{The integrity of a quantum-corrected black hole with ordinary or phantom global monopoles}

This kind of quantum-corrected AdS black holes with phantom global monopoles have complex structure, so we must scrutinize the whole metric to discuss the integrity. We have to elaborate the influences from the structure factors one by one on the entropy difference between the two stages for the black holes. We start to study the entropy of quantum-corrected black hole with two kinds of global monopoles in the world without specific cosmological constant. The metric of the black holes can be written as [19-21],
\begin{align}
	ds^{2}=g_{\mu\nu}dx^{\mu}dx^{\nu}\hspace{3cm}\notag\\
	=K(r)dt^{2}-\frac{dr^{2}}{K(r)}-r^{2}(d\theta^{2}+\sin^{2}\theta d\theta)
\end{align}
where
\begin{align}
	K(r)=\frac{\sqrt{r^{2}-\alpha^{2}}}{r}-\frac{2M}{r}-8\pi G\eta^{2}\xi
\end{align}
with Newton constant $G$ and $\alpha=4l_{p}$ with small magnitude showing the qnantum fluctuation. $l_{p}$ is the Planck length. As a monopole parameter, $\eta$ is of order $10^{16}GeV$ in a typical grand unified theory, leading to $8\pi G\eta^{2}\approx10^{-5}$ while $\xi=\pm1$ for regular global or phantom ones respectively [8-11]. $M$ is the mass parameter. The horizon of this kind of black holes as the root of $K(r)=0$ satisfies [19-21],
\begin{align}
	[1-(8\pi G\eta^{2}\xi)^{2}]r_{H}^{2}-32\pi G\eta^{2}\xi Mr_{H}-(4M^{2}+\alpha^{2})=0
\end{align}
and the horizon radius for black hole with metric (1) is [19-21],
\begin{align}
	r_{H}^{2}(M, \alpha, \eta^{2}, \xi)\hspace{4cm}\notag\\
	=\frac{1}{4[1-(8\pi\eta^{2}\xi)^{2}]^{2}}(4M+2|8\pi\eta^{2}\xi|\hspace{1cm}\notag\\
	\times[4M^{2}+\alpha^{2}(1-(8\pi\eta^{2}\xi)^{2})]^{\frac{1}{2}})^{2}
	+\alpha^{2}
\end{align}
The event horizon is a function of the variables due to rich structure like $r_{H}=r_{H}(M, \alpha, \eta^{2}, \xi)$. The Bekenstein-Hawking entropy of black hole is proportional to the horizon area [35-37],
\begin{align}
	S=\frac{1}{4}A_{H}
\end{align}
where
\begin{align}
	A_{H}=4\pi r_{H}^{2}
\end{align}

Here we assume that the quantum-corrected black hole involving global monopoles may break into two parts with the same structure. This kind of thermodynamic stability depends on the entropy difference of black holes in view of the second law of thermodynamics [29]. When the entropy decreases during the evolution of black hole, the black hole will be stable instead of splitting. In the process of the fragmentation, the original black hole can be thought as the initial state and the final state consists of two broken black holes under the conservation of mass. Before the separation, the entropy of the original black hole can be obtained according to Eq.(5) and Eq.(6),
\begin{align}
	S_{i}=\pi r_{H}^{2}(M, \alpha, \eta^{2}, \xi)
\end{align}
If the original black hole specified by the metric (1) is divided into two parts possessing the mass $\varepsilon_{M}M$ and $(1-\varepsilon_{M})M$ respectively, the divided black holes will have the metrics as follow,
\begin{align}
	ds^{2}=K_{i}(r)dt^{2}-\frac{dr^{2}}{K_{i}(r)}-r^{2}(d\theta^{2}+\sin^{2}\theta d\varphi^{2})
\end{align}
where $i=1, 2$ and the metric component functions are,
\begin{align}
	K_{1}(r)=\frac{\sqrt{r^{2}-\alpha^{2}}}{r}-\frac{2\varepsilon_{M}M}{r}-8\pi G\eta^{2}\xi
\end{align}
and
\begin{align}
	K_{2}(r)=\frac{\sqrt{r^{2}-\alpha^{2}}}{r}-\frac{2(1-\varepsilon_{M})M}{r}-8\pi G\eta^{2}\xi
\end{align}
The new horizon radii from Eq.(9) and (10) have their own expressions similar to Eq.(4) while the mass $M$ in Eq.(4) must be replaced as $\varepsilon_{M}M$ or $(1-\varepsilon_{M})M$. If the division of the isolated black hole happens, the entropy for the final state will be,
\begin{align}
	S_{f}=\pi r_{H}^{2}(\varepsilon_{M}M, \alpha, \eta^{2}, \xi)\notag\hspace{2cm}\\
	+\pi r_{H}^{2}((1-\varepsilon_{M})M, \alpha, \eta^{2}, \xi)
\end{align}
where $r_{H}(\varepsilon_{M}M, \alpha, \eta^{2}, \xi)$ and $r_{H}((1-\varepsilon_{M})M, \alpha, \eta^{2}, \xi)$ are the horizons of the fragmented black holes with masses $\varepsilon_{M}M$ and $(1-\varepsilon_{M})M$ respectively and also relate to the variables like $\alpha$, $\eta^{2}$, $\xi$ as description of black hole constitution. As mentioned in Ref.[32], we define the mass ratio and its region $0\leqslant\varepsilon_{M}\leqslant 1$. According to Eq.(7) and (11), the entropy difference can be written as [32],
\begin{align}
	\Delta S=S_{f}-S_{i}\notag\hspace{4cm}\\
	=[\pi r_{H}^{2}(\varepsilon_{M}M, \alpha, \eta^{2}, \xi)\notag\hspace{3cm}\\
	+\pi r_{H}^{2}((1-\varepsilon_{M})M, \alpha, \eta^{2}, \xi)]\hspace{1.5cm}\notag\\
	-\pi r_{H}^{2}(M, \alpha, \eta^{2}, \xi)\hspace{3.4cm}\notag\\
	=\frac{\pi}{4[1-(8\pi\eta^{2}\xi)^{2}]^{2}}\hspace{3.5cm}\notag\\
	\times\{-32\varepsilon(1-\varepsilon)M^{2}\hspace{3.2cm}\notag\\
	+16\varepsilon M|8\pi\eta^{2}\xi|\hspace{3.6cm}\notag\\
	\times[4\varepsilon^{2}M^{2}+\alpha^{2}(1-(8\pi\eta^{2}\xi)^{2}]^{\frac{1}{2}}\hspace{0.5cm}\notag\\
	+16(1-\varepsilon)M|8\pi\eta^{2}\xi|\hspace{2.8cm}\notag\\
	\times[4(1-\varepsilon)^{2}M^{2}+\alpha^{2}(1-(8\pi\eta^{2}\xi)^{2}]^{\frac{1}{2}}\notag\\
	-16M|8\pi\eta^{2}\xi|\hspace{3.8cm}\notag\\
	\times[4M^{2}+\alpha^{2}(1-(8\pi\eta^{2}\xi)^{2}]^{\frac{1}{2}}\hspace{1cm}\notag\\
	+4(8\pi\eta^{2}\xi)^{2}\hspace{4.2cm}\notag\\
	\times[-8\varepsilon(1-\varepsilon)M^{2}+\alpha^{2}(1-(8\pi\eta^{2}\xi))]\}\notag\\
	+\pi\alpha^{2}\hspace{5.2cm}
\end{align}
It is manifest that the entropy difference depends on various compositions of the black holes. It should be pointed out that the factor $\xi$ determining which kind of global monopoles appears in its square or absolute value form in the expression above, so whether $\xi=1$ or $\xi=-1$ will not change the entropy difference including its sign in this case. In the Figure 1, it is shown that the difference magnitude increases with stronger quantum gravity effect and the magnitude can become positive. The sufficiently strong effect may lead the black holes containing the global monopoles under the quantum fluctuation influence to break into two parts according to the second law of thermodynamics [29]. It is clear that the larger monopole factor reduces the value of entropy difference for the quantum-corrected black holes with regular or phantom global monopoles in the Figure 2. The global monopoles possessing greater energy scale will stabilize the compact bodies. In view of the curves in the Figure 1 and Figure 2, we infer that the black hole may become two parts, small black hole and large one.

\begin{figure}
	\centering
	\includegraphics[width=8cm]{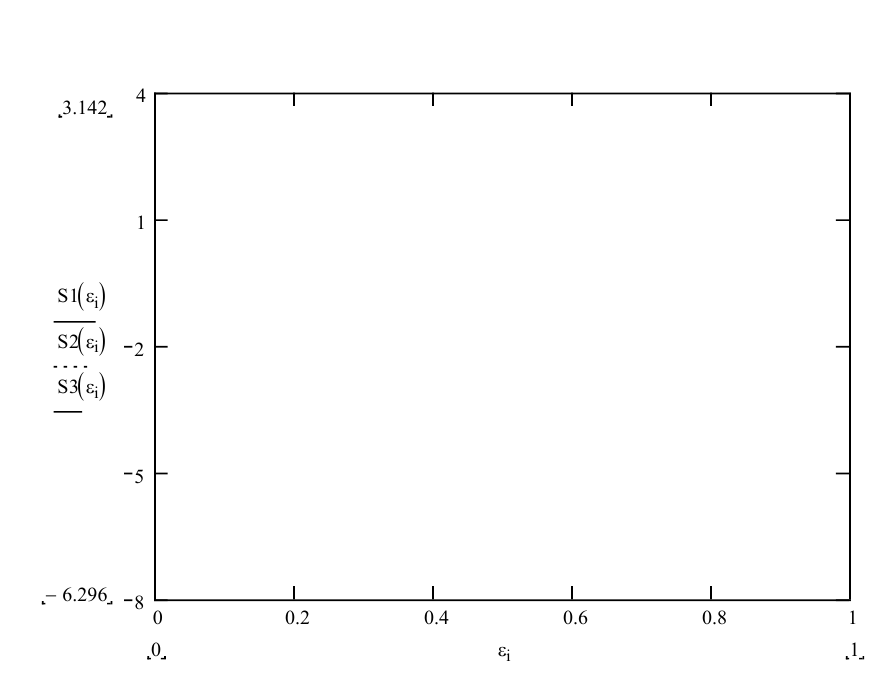}
	\caption{The solid, dotted and dashed curves of entropy difference for a quantum-corrected AdS black hole with ordinary or phantom global monopoles as functions of mass ratio $\varepsilon_{M}$ for $\alpha=0, 0.5, 1$ respectively while $8\pi G\eta^{2}=0.001$}
\end{figure}

\begin{figure}
	\centering
	\includegraphics[width=8cm]{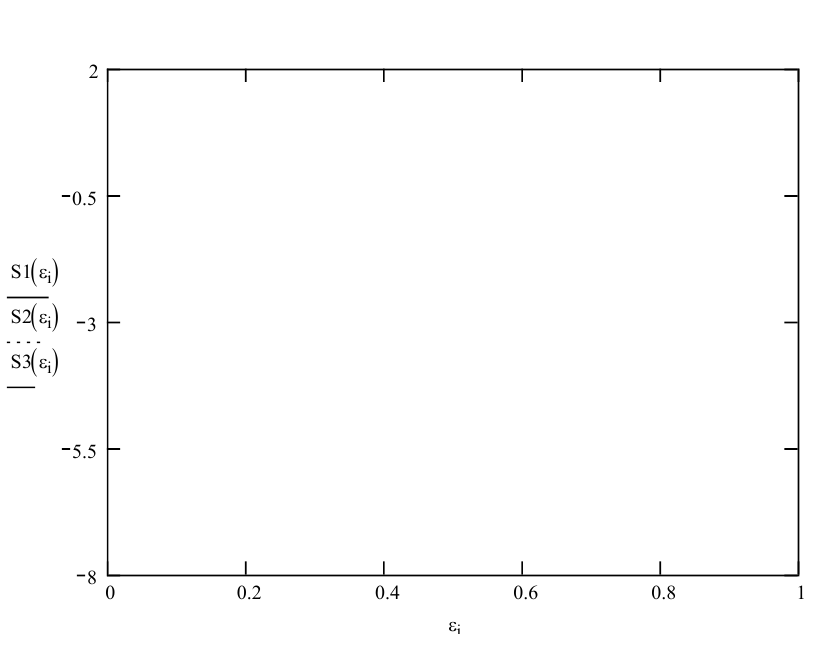}
	\caption{The solid, dotted and dashed curves of entropy difference a quantum-corrected AdS black hole with ordinary or phantom global monopoles as functions of mass ratio $\varepsilon_{M}$ for $8\pi G\eta^{2}=0.001, 0.05, 0.1$ respectively while $\alpha=0.1$}
\end{figure}

\section{The integrity of a quantum-corrected AdS black hole with ordinary or phantom global monopoles}

 Now we focus on the quantum-corrected black hoes swallowing the global monopoles within the anti-de Sitter spacetime. The spherically symmetric form of metric was found [19-21],
\begin{align}
	ds^{2}=g_{\mu\nu}dx^{\mu}dx^{\nu}\hspace{3cm}\notag\\
	=F(r)dt^{2}-\frac{dr^{2}}{F(r)}-r^{2}(d\theta^{2}+\sin^{2}\theta d\theta)
\end{align}
where
\begin{align}
	F(r)=\frac{\sqrt{r^{2}-\alpha^{2}}}{r}-\frac{2M}{r}-8\pi G\eta^{2}\xi+\frac{(r^{2}-\alpha^{2})^{\frac{3}{2}}}{rl^{2}}
\end{align}
where the source structure is characterized with various parameters mentioned previously. Here $\Lambda=-\frac{3}{l^{2}}$ is the cosmological constant for AdS universe. In order to consider the entropy, we  must find the root of $F(r)=0$ from component function (14) to obtain the event horizon of black hole like,
\begin{align}
	(r^{2}-\alpha^{2})^{3}+2l^{2}(r^{2}-\alpha^{2})^{2}-4Ml^{2}(r^{2}-\alpha^{2})^{\frac{3}{2}}\notag\\
	+[l^{2}-(8\pi G\eta^{2}\xi l^{2})^{2}](r^{2}-\alpha^{2})\notag\hspace{2cm}\\
	-4Ml^{4}(r^{2}-\alpha^{2})^{\frac{1}{2}}\notag\hspace{3cm}\\
	+[4(Ml^{2})^{2}-(8\pi G\eta^{2}\xi l^{2}\alpha)^{2}]=0
\end{align}
This is a sixth-degree equation for $\sqrt{r^{2}-\alpha^{2}}$. The event horizon can be thought as function of variables due to rich structure, $r_{H}=r_{H}(M, \alpha, \eta^{2}, \xi, l)$. It is found that the factor $\xi$ for the global monopole species appears in the square form in Eq.(15). The entropy of the original black hole can be obtained according to Eq.(5) and (6),
\begin{align}
	S_{i}=\pi r_{H}^{2}(M, \alpha, \eta^{2}, \xi, l)
\end{align}
If the original black hole with the description of the metric (13) is divided into two parts possessing the mass $\varepsilon_{M}M$ and $(1-\varepsilon_{M})M$ respectively, the divided black holes will have the metrics as follow,
\begin{align}
	ds^{2}=F_{i}(r)dt^{2}-\frac{dr^{2}}{F_{i}(r)}-r^{2}(d\theta^{2}+\sin^{2}\theta d\varphi^{2})
\end{align}
where $i=1, 2$ and the metric component functions are,
\begin{align}
	F_{1}(r)=\frac{\sqrt{r^{2}-\alpha^{2}}}{r}-\frac{2\varepsilon_{M}M}{r}-8\pi G\eta^{2}\xi+\frac{(r^{2}-\alpha^{2})^{\frac{3}{2}}}{l^{2}r}
\end{align}
and
\begin{align}
	F_{2}(r)=\frac{\sqrt{r^{2}-\alpha^{2}}}{r}-\frac{2(1-\varepsilon_{M})M}{r}-8\pi G\eta^{2}\xi\notag\\
	+\frac{(r^{2}-\alpha^{2})^{\frac{3}{2}}}{l^{2}r}\hspace{3cm}
\end{align}
The new horizon radii from Eq.(8) and (9) satisfy the equations similar to Eq.(3) while the mass $M$ in Eq.(3) must be replaced as $\varepsilon_{M}M$ or $(1-\varepsilon_{M})M$. The revised sixth-degree equations for $\sqrt{r^{2}-\alpha^{2}}$ have roots $r_{H}(\varepsilon_{M}M, \alpha, \eta^{2}, \xi, l)$ and $r_{H}((1-\varepsilon_{M})M, \alpha, \eta^{2}, \xi, l)$. If the division of the isolated black hole happens, the entropy for the final state will be,
\begin{align}
	S_{f}=\pi r_{H}^{2}(\varepsilon_{M}M, \alpha, \eta^{2}, \xi, l)\notag\hspace{2cm}\\
	+\pi r_{H}^{2}((1-\varepsilon_{M})M, \alpha, \eta^{2}, \xi, l)
\end{align}
where $r_{H}(\varepsilon_{M}M, \alpha, \eta^{2}, \xi, l)$ and $r_{H}((1-\varepsilon_{M})M, \alpha, \eta^{2}, \xi, l)$ are the horizons of the fragmented black holes with masses $\varepsilon_{M}M$ and $(1-\varepsilon_{M})M$ respectively and also relate to the variables like $\alpha$, $\eta^{2}$, $\xi$, $l$ as description of black hole constitution. As mentioned in Ref.[32], we define the mass ratio and its region $0\leqslant\varepsilon_{M}\leqslant 1$. According to Eq.(6) and (10), the entropy difference can be written as [32],
\begin{align}
	\Delta S=S_{f}-S_{i}\notag\hspace{3.5cm}\\
	=[\pi r_{H}^{2}(\varepsilon_{M}M, \alpha, \eta^{2}, \xi, l)\notag\hspace{2cm}\\
	+\pi r_{H}^{2}((1-\varepsilon_{M})M, \alpha, \eta^{2}, \xi, l)]\notag\\
	-\pi r_{H}^{2}(M, \alpha, \eta^{2}, \xi, l)\hspace{2cm}
\end{align}
The difference certainly has something to do with the structure factors. The sign of the entropy difference determines whether the division of the isolated black hole can appear spontaneously based on the second law of thermodynamics under the influences from quantum fluctuation, global monopole and the AdS radius [29]. The negative nature of the difference shows that the black holes are stable instead of rupturing. The positive ones indicates that the compact bodies may split. It is more important to explore how the structure affects the evolution of black hole and further whether the total effect may change the sign of the difference. It is difficult to find the event horizon as solutions to the sixth-degree equation (15). The solutions to Eq.(15) are certainly functions of several variables, but the approximate numerical solutions can not help us to manifest the dependence of entropy difference (23) on the variables perfectly.

Having analyzed the metric of a quantum-corrected AdS black holes with regular or phantom global monopoles in Eq.(13) and (14), we discover that there is no coupling term for monopole scale $\eta$ and radius of AdS universe $l$. It is  useful to investigate how the cosmological constant modifies the existence of the gravitational source in the AdS universe. Without loss of generality, it is acceptable to ignore the term with $\eta$ temporarily. In the research on the effect from the cosmological scale, we consider the metric with factor $\alpha$ [22],
\begin{align}
	ds^{2}=(\frac{\sqrt{r^{2}-\alpha^{2}}}{r}-\frac{2M}{r}+\frac{(r^{2}-\alpha^{2})^{\frac{3}{2}}}{rl^{2}})dt^{2}\notag\\
	-\frac{dr^{2}}{\frac{\sqrt{r^{2}-\alpha^{2}}}{r}-\frac{2M}{r}+\frac{(r^{2}-\alpha^{2})^{\frac{3}{2}}}{rl^{2}}}\notag\hspace{1cm}\\
	-r^{2}(d\theta^{2}+\sin^{2}\theta d\varphi^{2})\hspace{2cm}
\end{align}
because of the last term as coipling of quantum fluctuation and the cosmological radius. In view of the metric (22), the Eq.(15) for the horizon is reduced to be,
\begin{align}
	(\sqrt{r^{2}-\alpha^{2}})^{3}+l\sqrt{r^{2}-\alpha^{2}}-2Ml^{2}=0
\end{align}
The horizon radius as solution to Eq.(23),
\begin{align}
	r_{H}(M, \alpha, l)=[(Ml^{2}+\sqrt{M^{2}l^{4}+\frac{l^{3}}{27}})^{\frac{2}{3}}\notag\hspace{1cm}\\
	+(Ml^{2}-\sqrt{M^{2}l^{4}+\frac{l^{3}}{27}})^{\frac{2}{3}}\notag\\
	-\frac{2}{3}l+\alpha^{2}]^{\frac{1}{2}}\hspace{2cm}
\end{align}
We can obtain the difference that entropy of the final black hole minus the initial black hole's one,
\begin{align}
	\Delta S=S_{f}-S_{i}\notag\hspace{4.3cm}\\
	=[\pi r_{H}^{2}(\varepsilon_{M}M, \alpha, l)+\pi r_{H}^{2}((1-\varepsilon_{M})M, \alpha, l)]\notag\\
	-\pi r_{H}^{2}(M, \alpha, l)\notag\hspace{3.9cm}\\
	=\pi(\varepsilon_{M}Ml^{2}+\sqrt{\varepsilon_{M}^{2}M^{2}l^{4}+\frac{l^{3}}{27}})^{\frac{2}{3}}\notag\hspace{1.5cm}\\
	+\pi(\varepsilon_{M}Ml^{2}-\sqrt{\varepsilon_{M}^{2}M^{2}l^{4}+\frac{l^{3}}{27}})^{\frac{2}{3}}\notag\hspace{1cm}\\
	+\pi[(1-\varepsilon_{M})Ml^{2}+\sqrt{\varepsilon_{M}^{2}M^{2}l^{4}+\frac{l^{3}}{27}}]^{\frac{2}{3}}\notag\\
	+\pi[(1-\varepsilon_{M})Ml^{2}-\sqrt{\varepsilon_{M}^{2}M^{2}l^{4}+\frac{l^{3}}{27}}]^{\frac{2}{3}}\notag\\
	-\pi(Ml^{2}+\sqrt{M^{2}l^{4}\frac{l^{3}}{27}})^{\frac{2}{3}}\notag\hspace{2cm}\\
	-\pi(Ml^{2}-\sqrt{M^{2}l^{4}\frac{l^{3}}{27}})^{\frac{2}{3}}\notag\hspace{2cm}\\
	-\frac{2}{3}\pi l+\pi\alpha^{2}\hspace{3.9cm}
\end{align}

The dependence of entropy difference on the division ratio is depicted in the pictures. The values of entropy difference denoted as the leftmost and rightmost parts of profiles are positive in the Figure 3. The profiles from Figure 3 show that the fluctuation may lead the difference to be larger than the zero while the ratio is tiny or approaches the unit. The AdS scale modifies the difference curves. The larger scale enhances the value of entropy change associated with the mid-part of ratio. It is obvious that the larger cosmological scale relating to the cosmological constant encourages the difference value to be positive within the mid-region for ratio according to Figure 4. It is interesting that the stronger fluctuation increases the magnitude of the whole change in entropy between the two stages for the black hole, which is consistent with the results from Eq.(16). The stability whether indivisibility or fragmentation of quantum-corrected AdS black hole swallowing regular or phantom monopoles depends on the final effects from black hole structure with several factors. The quantum fluctuation and the cosmological scale of AdS background can increase the fragmentation possibility of black hole. After the division, one is extremely small and the other is giant between the two parts.

\begin{figure}
	\centering
	\includegraphics[width=8cm]{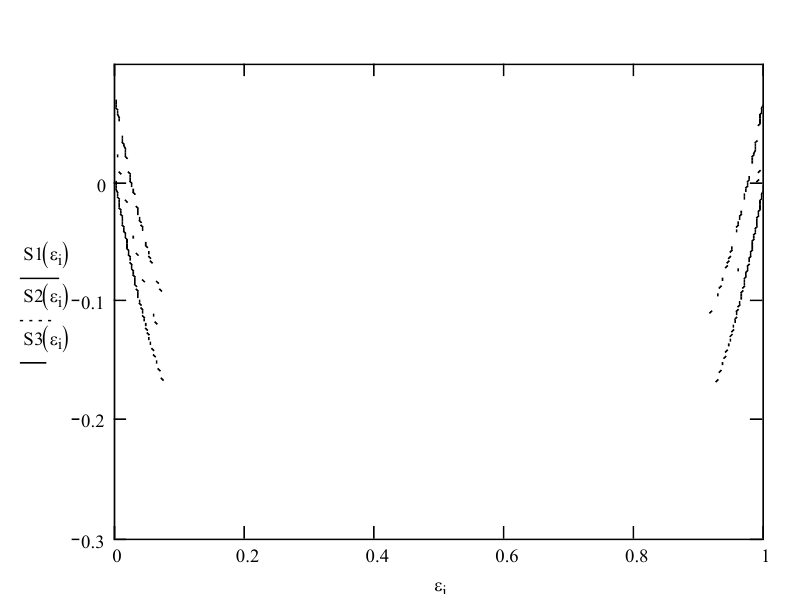}
	\caption{The solid, dotted and dashed curves of entropy difference as functions of mass ratio $\varepsilon_{M}$ for $\alpha=0, 0.1, 0.15$ respectively while $l=1$}
\end{figure}

\begin{figure}
	\centering
	\includegraphics[width=8cm]{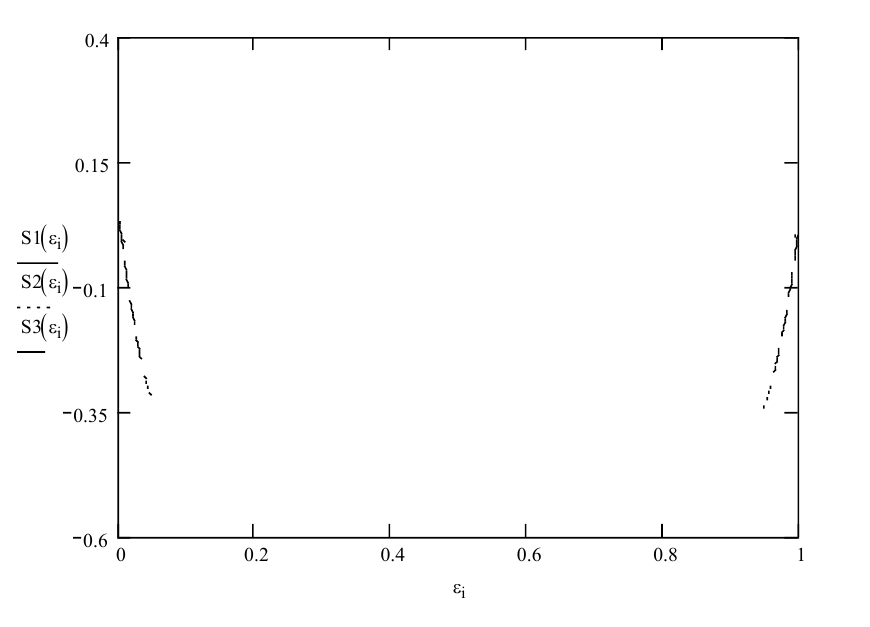}
	\caption{The solid, dotted and dashed curves of entropy difference as functions of mass ratio $\varepsilon_{M}$ for $l=1, 1.5, 2.5$ respectively while $\alpha=0.1$}
\end{figure}

\section{Conclusion}

We scrutinize the fragmentation stability of the quantum-corrected AdS black holes with phantom global monopoles. More efforts have been contributed to the black holes [22]. The thermodynamic quantities of the black holes such as the Hawking temperature, entropy, heat capacity were calculated and the geodesic equations, shadow and energy emission rate for this kind of black holes were investigated [22]. The thermodynamic stability was also studied in view of the nature of heat capacity relating to the black hole structures and the black holes evaporate to disappear with negative heat capacity [22]. The fragmentation of black holes can violate their indivisibility also thought as another thermodynamic stability. We explore the possibility that the black holes with complicated structure may break into two parts under the second law of thermodynamics. The quantum corrections and the AdS background may stimulate the separation of the black hole. We derive the entropies of the isolated black hole and the fragmented parts respectively and estimate the entropy difference between the stages, the initial black hole for the first stage and the two broken parts in the final case. We discuss the sign of the entropy difference as a function of monopole parameter, fluctuation factor and the AdS world size. The influence from the regular monopole or phantom one is weaker on the nature of the difference. The considerable fluctuation from quantum gravity may encourage the black hole to split, one tiny black hole and other huge one. The AdS scale can increase the magnitude of the whole entropy difference and the large enough AdS radius may let the initial black hole to be divided nearly half-and-half. We can further the characteristics.

\vspace{1cm}
\noindent \textbf{Acknowledge}

This work is partly supported by the Shanghai Key Laboratory of
Astrophysics 18DZ2271600.

\end{document}